\documentclass[aps,prl,reprint,superscriptaddress]{revtex4-2}
\usepackage{amsthm,amsfonts,amsmath,amssymb}
\usepackage{mathrsfs}
\usepackage{physics}
\usepackage{autobreak}
\usepackage{textcomp}
\usepackage{hyperref}
\usepackage{graphicx}
\usepackage{float}

\begin{document}
\title{Instantaneous and Retarded Interactions in Coherent Radiation }
\author{Zhuoyuan Liu}
\affiliation{Department of Engineering Physics, Tsinghua University, Beijing 100084, China}
\author{Xiujie Deng}
\affiliation{Institute for Advanced Study, Tsinghua University, Beijing 100084, China}
\author{Tong Li}
\affiliation{Department of Engineering Physics, Tsinghua University, Beijing 100084, China}
\author{Lixin Yan}
\email[]{yanlx@mail.tsinghua.edu.cn}
\affiliation{Department of Engineering Physics, Tsinghua University, Beijing 100084, China}

\date{\today}

\begin{abstract}
In coherent radiation of an ensemble of electrons, radiation field from electrons
resonantly drives the other electrons inside to produce stimulated emission. The radiation
reaction force on the electrons accounting for this stimulated radiation loss 
is classically described by the Li\'{e}nard-Wiechert potential. Despite its being the foundation
of beam physics for decades, we show that using the ``acceleration field'' in Li\'{e}nard-Wiechert 
potential to describe radiative interactions leads to divergences due to its implicit dependence on 
instantaneous interactions. Here, we propose an alternative theory for electromagnetic radiation 
by decomposing the interactions into instantaneous part and retarded part. It is shown that
only the retarded part contributes to the irreversible radiation loss and the instantaneous 
part describes the space charge related effects. We further apply this theory to study the coherent 
synchrotron radiation wake, which hopefully will reshape our understanding of coherent radiation
and collective interactions.
\end{abstract}

\maketitle

\paragraph{Introduction.---}
The retarded solutions of Maxwell equations form the foundation of classical electrodynamics
(CED). In particular, the retarded potential of point charges, known as the
Li\'{e}nard-Wiechert potential\cite{jackson1998classical},
which connects the particle's past trajectory to the present field
observed, has been widely applied across photonics, beam physics and astrophysics. 

In most context, the two terms in Li\'{e}nard-Wiechert solution, the ``velocity field''
and the ``acceleration field'', are attributed to the name ``space charge field" and 
``radiation field" for their $\mathcal{O}(1/R^2)$ and $\mathcal{O}(1/R)$ 
behavior of decay and dependence on acceleration. However, such an interpretation 
only works in the limit of acceleration $\dot{\boldsymbol{\beta}}\rightarrow 0$
or observation distance $R\rightarrow\infty$. For the general case, both the ``velocity
field" and ``acceleration field'' are a mixture of contributions from space charge and
radiation, and their physical meaning become obscure. Therefore, such a classification
of the Li\'{e}nard-Wiechert solution is often misleading in analysis of interactions between 
charged particles.

In the past decades, this retarded solution has been largely applied to the study
of the collective radiative interactions in high-current and low emittance 
electron beams in modern accelerator facilities, known as the coherent synchrotron 
radiation (CSR) effect. Due to the collective enhancement of radiation, free electron
lasers and storage rings have utilized CSR mechanism to generate radiation with 
unprecedented brightness\cite{franz2024terawatt,evain2019stable}. Meanwhile, CSR also 
induces prominent energy spread and emittance growth detrimental to further 
applications. Early attempts to evaluate 1D CSR wakefield \cite{saldin1997on,mayes2009exact} 
involve artificial renormalization techniques to eliminate singularities and extract 
radiation reaction (RR) force from the retarded solution. 2D or 3D CSR theory has 
been proved to have finite wakefield. However, these attempts start from the 
``acceleration field'' which does not correspond to RR force\cite{cai2017coherent,huang2013two}, 
thus conflicting with the renormalized 1D theory. This inconsistency among different 
formulations of CSR theory reflects the misconceptions about Li\'{e}nard-Wiechert 
potentials and a lack of understanding of RR force in coherent radiation.

Some pioneering work has devoted to the study of RR force in coherent radiation, but
is only achieved by phenomenologically ensuring energy conservation \cite{kimel1995coherent}
or analyzing some toy models using other formulations of CED \cite{niknejadi2015radiated}
which lacks of practicality. To the best of our knowledge, there is no explicit and 
comprehensive formulation for coherent RR force derived from first principles.

In this letter, we adopt a different physical picture to describe electromagnetic
radiation of moving charges. We analytically derive the coherent RR force
by decomposing the fields into instantaneous and retarded parts. The work of 
the retarded field on electrons gives the dissipative radiation loss. The
instantaneous field, on the contrary, describes the momentum exchange between
electrons, which can be further decomposed to the classical space 
charge field and ``compression field" proposed by Dohlus\cite{dohlus2000coherent}.
Finally, we calculate the steady-state 2D CSR wake of a circular orbit, numerically
validating our theory.
In this way, we reconcile the historical action-at-a-distance Newtonian dynamics and
Farady-Maxwell retarded electrodynamics, and derive a self-consistent theory describing
coherent radiation.

\paragraph{Decomposition of fields.---}
We start from the non-homogeneous Maxwell equation for electric field
\begin{equation}
    \laplacian{\boldsymbol{E}} - \frac{1}{c^2}\pdv[2]{\boldsymbol{E}}{t} = 
    \frac{1}{\epsilon_0}\left(\grad{\rho}+\frac{1}{c^2}\pdv{\boldsymbol{J}}{t}\right),
\end{equation}
where ${\rho}$ is the charge density and ${\boldsymbol{J}}$ is the current density. 
Let us first consider the case of a point charge with ${\rho}=e\delta(\boldsymbol{r}-\boldsymbol{r}_0(t))$
and $\boldsymbol{J} = ec\boldsymbol{\beta}(t)\delta(\boldsymbol{r}-\boldsymbol{r}_0(t))$.
In space-time Fourier domain the solution is
\begin{equation}
    \tilde{\boldsymbol{E}}(\boldsymbol{k},\omega)=
    \frac{i\boldsymbol{k}c^2\tilde{\rho}(\boldsymbol{k},\omega)-i\omega\tilde{\boldsymbol{J}}(\boldsymbol{k},\omega)}
	{\epsilon_0 (\omega^2-k^2c^2)}.
    \label{eq2}
\end{equation}

We seek a decomposition of the field into instantaneous fields including 
space charge effects and retarded field describing RR.
Since radiative interaction is mediated by real photons which are transverse, 
the RR field is itself transverse, which means its propagation direction is
transverse to its polarization. Thus, we decompose
the solution of Maxwell equation into transverse and longitudinal component.
The longitudinal electric field is the Coulomb field generated by the 
instantaneous charge density, which is consistent with our claim that the
radiative interactions are included in the retarded transverse field.
\begin{equation}
    \boldsymbol{E}^{\parallel}(\boldsymbol{r},t) = 
    \frac{e}{4\pi\epsilon_0}
    \frac{\boldsymbol{r}_0(t)-\boldsymbol{r}}{|\boldsymbol{r}_0(t)-\boldsymbol{r}|^3}.
    \label{eq3}
\end{equation}

\begin{figure}[t]
    \includegraphics[width=200pt]{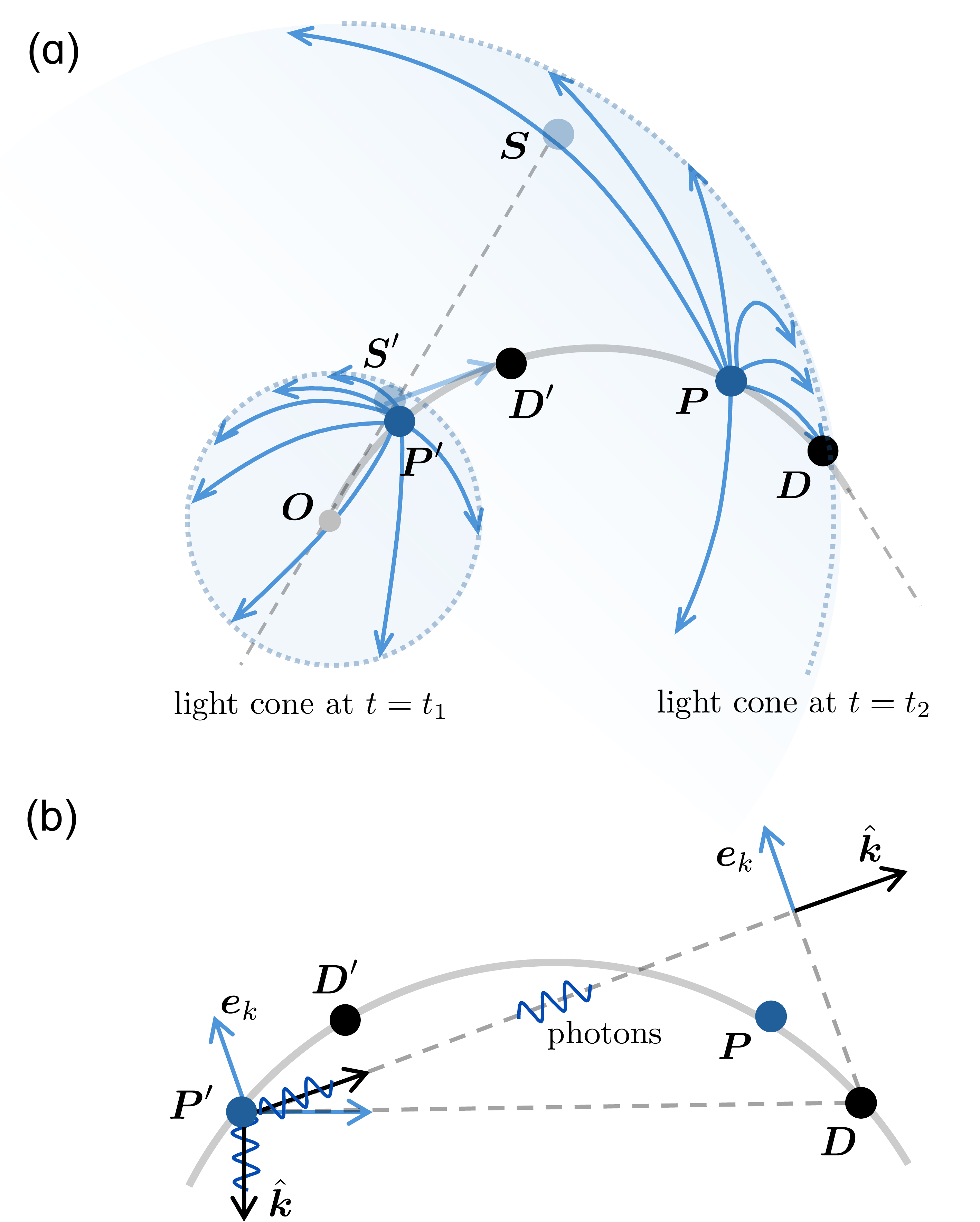}
    \caption{Schematic of different perspectives on radiative interaction of
    two point charges. (a) In the classical Li\'{e}nard-Wiechert potential 
    interpretation, the field on the test particle $\boldsymbol{D}^\prime$
    is the Coulomb field of the virtual charge $\boldsymbol{S}^\prime$
    moving along the tangent direction to the entrance of the arc $\boldsymbol{O}$ if the
    light cone (the bubble around $\boldsymbol{P}^\prime$ centered $\boldsymbol{O}$
    in dotted line) does not catch up with $\boldsymbol{D}^\prime$, 
    which is coined the entrance transient. For steady state interaction, 
    the light cone from $\boldsymbol{O}$ encloses the test particle 
    $\boldsymbol{D}$ and the field on $\boldsymbol{D}$ is determined by the 
    retarded source point $\boldsymbol{O}$. (b) In our model,
    the field is first decomposed to propagate along different directions.
    The field with polarization vector parallel to the direction
    $\overline{\boldsymbol{P}^\prime\boldsymbol{D}}$ will instantaneously
    interact with the test particle at $\boldsymbol{D}$, while the other
    fields will interact with $\boldsymbol{D}$ in a retarded way.}
    \label{fig1}
\end{figure}

For the transverse electric field, only the transverse current density term 
needs to be retained. We assume a complete set of basis for $\mathbb{R}^3$, 
namely, $(\boldsymbol{e}_{k1},\boldsymbol{e}_{k2},\hat{\boldsymbol{k}})$, 
where $\hat{\boldsymbol{k}}$ is the unit vector along the wave vector
$\boldsymbol{k}$, and the two polarization vectors 
$\boldsymbol{e}_{k\nu}$ are transverse to $\hat{\boldsymbol{k}}$. 
We project Eq.\eqref{eq2} to the directions of polarization and make an
inverse temporal Fourier transform along the retarded contour 
$\omega\pm kc \rightarrow \omega\pm kc-i0^+$, obtaining
\begin{equation}
    \boldsymbol{E}^{\bot} = -\frac{ec}{2\epsilon_0}
    \int_{-\infty}^{+\infty}k^2 \dd{k}\int\frac{\dd{\Omega}}{(2\pi)^3}
    \int_{-\infty}^{t}\dd{t^\prime}
    \boldsymbol{\beta}_\bot e^{ik\phi},
    \label{eq4}
\end{equation}
where we define the transverse velocity $\boldsymbol{\beta}_\bot=
\sum_\nu (\boldsymbol{\beta}\cdot \boldsymbol{e}_{k\nu})\boldsymbol{e}_{k\nu}$ 
and the retardation
$\phi=\hat{\boldsymbol{k}}\cdot(\boldsymbol{r}-\boldsymbol{r}_0(t^\prime))
-c(t-t^\prime)$.

The usual retarded solution is obtained by first integrating
over solid angle and then the wave number. This, along with the instantaneous
Coulomb field, gives the electric field fully determined by the motion of 
charges at retarded time. This expression of field in variables of dynamics 
at retarded time leads to the Li\'{e}nard-Wiechert solution and 
thereafter the classical picture of CSR, which is shown in 
Fig.\ref{fig1}(a). For two particles moving along a circular orbit, the 
test particle ahead only sees the fields generated by the source particle 
at retarded position.

However, if we view the radiative interaction in angular domain,
there are forces acting instantaneously on the observer due to the fields 
propagating perpendicular to the direction of observation, as shown in 
Fig.\ref{fig1}(b). This indicates that although the Li\'{e}nard-Wiechert 
solution is fully determined at retarded time, it still implicitly 
involves instantaneous interactions which is the reason why we encountered
diverging force in describing radiative interaction.
In this regard, to discriminate the contributions from instantaneous and retarded fields,
we first integrate Eq.\eqref{eq4} over wave number and subsequently integrate
over past time, resulting in the remaining integration over solid angle
\begin{equation}
    \begin{aligned}
        \boldsymbol{E}^{\bot} = &
        \frac{ec}{2\epsilon_0}\int\frac{\dd{\Omega}}{(2\pi)^2}
        \int_{-\infty}^{t}\dd{{t^\prime}} \boldsymbol{\beta}_\bot
        \delta^{\prime\prime}(\phi)\\
        = & \frac{ec}{2\epsilon_0}\int\frac{\dd{\Omega}}{(2\pi)^2}
        \left[\boldsymbol{\mathcal{B}}_{1}\,\delta^\prime(\phi)
        -\boldsymbol{\mathcal{B}}_{2}\,\delta(\phi)\right]_{t^\prime=t}\\
        &+ \frac{ec}{2\epsilon_0}\int\frac{\dd{\Omega}}{(2\pi)^2}
        \int_{-\infty}^{t}\boldsymbol{\mathcal{B}}_{3}\delta(\phi)\dd{t^\prime}.
    \end{aligned}
    \label{eq5}
\end{equation}
where we defined the following quantities
$\boldsymbol{\mathcal{B}}_{1} = \boldsymbol{\beta}_\bot \pdv{t^\prime}{\phi}$,
$\boldsymbol{\mathcal{B}}_{2} = \pdv{t^\prime}(\boldsymbol{\beta}_\bot \pdv{t^\prime}{\phi})\pdv{t^\prime}{\phi}$,
$\boldsymbol{\mathcal{B}}_{3} = \pdv{t^\prime}\left(\pdv{t^\prime}\left(\boldsymbol{\beta}_\bot\pdv{t^\prime}{\phi}\right)\pdv{t^\prime}{\phi}\right)$.

The first two terms of Eq.\eqref{eq5} are the instantaneous transverse fields 
which are singular due to its dependence on Dirac delta in the integrand. 
The last term is the retarded transverse field, which we call the RR field.
The RR field is everywhere finite since the singular Dirac delta function has
been integrated by the past time. 

The integration over solid angle in Eq.\eqref{eq5}
can be explicitly evaluated by assuming a frame where
$\boldsymbol{e}_{k1} = (\cos\theta\cos\varphi,\cos\theta\sin\varphi,-\sin\theta)$,
$\boldsymbol{e}_{k2} = (-\sin\varphi,\cos\varphi,0)$,
$\hat{\boldsymbol{k}} = (\sin\theta\cos\varphi,\sin\theta\sin\varphi,\cos\theta)$,
$\boldsymbol{\beta} = \beta(0,\sin\theta_b,\cos\theta_b)$ and
$\dv{\boldsymbol{\beta}}{t} = |\dv{\boldsymbol{\beta}}{t}|
(\sin\theta_a\cos\varphi_a,\sin\theta_a\sin\varphi_a,\cos\theta_a)$.
Here, we further simplify the problem by assuming $\boldsymbol{\beta}\bot
\dot{\boldsymbol{\beta}}$ which is satisfied for synchrotron radiation.
Combining with the result of instantaneous longitudinal electric
field, the total electric field can be decomposed into three terms,
the instantaneous velocity field $\boldsymbol{E}_\mathrm{vel}^\mathrm{inst}$,
the instantaneous acceleration field $\boldsymbol{E}_\mathrm{acc}^\mathrm{inst}$
and the radiation reaction field $\boldsymbol{E}_\mathrm{rr}$
\begin{equation}
    \boldsymbol{E} = \boldsymbol{E}_\mathrm{vel}^\mathrm{inst} +
    \boldsymbol{E}_\mathrm{acc}^\mathrm{inst} + \boldsymbol{E}_\mathrm{rr},
\end{equation}
with
\begin{equation}
    \begin{aligned}
        \boldsymbol{E}_\mathrm{vel}^\mathrm{inst} = &\frac{e}{4\pi\epsilon_0}
        \frac{\hat{\boldsymbol{R}}}{R^2}
        \frac{1-\beta^2}{(1-(\boldsymbol{\beta}\times\hat{\boldsymbol{R}})^2)^{3/2}},\\
        \boldsymbol{E}_\mathrm{acc}^\mathrm{inst} = &- \frac{e}{4\pi\epsilon_0}
        \frac{1}{cR}\frac{\dot{\boldsymbol{\beta}} + 
        \left(\dot{\boldsymbol{\beta}}\cdot\hat{\boldsymbol{R}}\right)
        \hat{\boldsymbol{R}}}{2(1-(\boldsymbol{\beta}\times\hat{\boldsymbol{R}})^2)^{3/2}}\\
        & + \frac{e}{4\pi\epsilon_0} \frac{1}{cR}
        \frac{3(\boldsymbol{\beta}\cdot\hat{\boldsymbol{R}})^2
        \left(\dot{\boldsymbol{\beta}}\cdot \hat{\boldsymbol{R}}\right)
        \hat{\boldsymbol{R}}}{2(1-(\boldsymbol{\beta}\times\hat{\boldsymbol{R}})^2)^{5/2}}.
    \end{aligned}
\end{equation}
where $\boldsymbol{R} = \boldsymbol{r}-\boldsymbol{r}_0(t)$, $\hat{\boldsymbol{R}} = \boldsymbol{R}/R$, 
$\dot{\boldsymbol{\beta}}=\dd{\boldsymbol{\beta}}/\dd{t}$. 
The explicit form of radiation reaction field $\boldsymbol{E}_{rr}$ can be 
found using the Li\'{e}nard-Wiechert solution for $\boldsymbol{E}$.

We can also decompose the magnetic field of a point charge in the same way.
As the magnetic field is a transverse field, the solution of Maxwell equation
\begin{equation}
    \nabla^2\boldsymbol{B} -\frac{1}{c^2}\pdv[2]{\boldsymbol{B}}{t} = -\mu_0 \curl{\boldsymbol{J}}
\end{equation}
can be directly written as
\begin{equation}
    \boldsymbol{B} = \frac{e}{2\epsilon_0}\int_{-\infty}^{+\infty}k^2\dd{k}
    \int\frac{\dd{\Omega}}{(2\pi)^3}\int_{-\infty}^{t}\dd{t^\prime} 
    \hat{\boldsymbol{k}}\times \boldsymbol{\beta}
    e^{ik\phi}.
\end{equation}

Following the lines of the decomposition of electric field, we have
the decomposition of magnetic field into instantaneous velocity field $\boldsymbol{B}_\mathrm{vel}^\mathrm{inst}$,
instantaneous acceleration field $\boldsymbol{B}_\mathrm{acc}^\mathrm{inst}$
and radiation reaction field $\boldsymbol{B}_\mathrm{rr}$
\begin{equation}
    \boldsymbol{B} = \boldsymbol{B}_\mathrm{vel}^\mathrm{inst}
    + \boldsymbol{B}_\mathrm{acc}^\mathrm{inst} + \boldsymbol{B}_\mathrm{rr}
\end{equation}
with
\begin{equation}
    \boldsymbol{B}_\mathrm{vel}^\mathrm{inst} =
    \frac{e}{4\pi\epsilon_0}\frac{1}{cR^2}
    \frac{(1-\beta^2)(\boldsymbol{\beta}\times\hat{\boldsymbol{R}})}
    {(1-(\boldsymbol{\beta}\times\hat{\boldsymbol{R}})^2)^{3/2}},
\end{equation}
\begin{widetext}
\begin{equation}
    \begin{aligned}
        \boldsymbol{B}_\mathrm{acc}^\mathrm{inst} = 
        \frac{e}{4\pi\epsilon_0}\left\{\frac{1}{c^2R}\frac{(\boldsymbol{\beta}\times\dot{\boldsymbol{\beta}})
        +(\hat{\boldsymbol{R}}\cdot\dot{\boldsymbol{\beta}})(\boldsymbol{\beta}\times\hat{\boldsymbol{R}})
        -2(\boldsymbol{\beta}\cdot\hat{\boldsymbol{R}})(\hat{\boldsymbol{R}}\times\dot{\boldsymbol{\beta}})}
        {2(1-(\boldsymbol{\beta}\times\hat{\boldsymbol{R}})^2)^{3/2}}
        - \frac{1}{c^2 R}
        \frac{3(\boldsymbol{\beta}\cdot\hat{\boldsymbol{R}})^2(\hat{\boldsymbol{R}}\cdot\dot{\boldsymbol{\beta}})
        (\boldsymbol{\beta}\times\hat{\boldsymbol{R}})}
        {2(1-(\boldsymbol{\beta}\times\hat{\boldsymbol{R}})^2)^{5/2}}\right\}.
    \end{aligned}
\end{equation}
\end{widetext}

\paragraph{Lorentz force.---}
The Lorentz force exerted on a test particle from the source particle can
be decomposed accordingly using the field decomposition
\begin{equation}
    \begin{aligned}
        \boldsymbol{F} = &\, \boldsymbol{F}_\mathrm{sc} + \boldsymbol{F}_\mathrm{comp}
        + \boldsymbol{F}_\mathrm{rr} \\
        = &\, e(\boldsymbol{E}_\mathrm{vel}^\mathrm{inst}+c\boldsymbol{\beta}\times
        \boldsymbol{B}_\mathrm{vel}^\mathrm{inst}) +
        e(\boldsymbol{E}_\mathrm{acc}^\mathrm{inst}+c\boldsymbol{\beta}\times
        \boldsymbol{B}_\mathrm{acc}^\mathrm{inst}) \\
        & +e(\boldsymbol{E}_\mathrm{rr}+c\boldsymbol{\beta}\times
        \boldsymbol{B}_\mathrm{rr}).
    \end{aligned}
\end{equation}
The first two terms are instantaneous, indicating that they are Newtonian
and obey the third law of Newtonian dynamics, i.e., the interaction
conserves the energy of the system. The first term remains when there is
no acceleration and coincides with the instantaneous electromagnetic field
derived from Li\'{e}nard-Wiechert potential, thus is identified as the
space charge force. The second term only exists if the charge distribution
of the beam changes, hence can be referred to the ``compression force''
defined by Dohlus\cite{dohlus2000coherent}. The last RR term is evaluated at retarded time,
representing the loss of kinetic energy through escaping radiation.

\paragraph{Poynting's theorem.---}
Having constructed a theory describing the radiative interaction and 
Newtonian interactions in coherent radiation, here we give a brief proof 
of the claim that only the retarded RR field accounts for the radiation 
loss into free space. In the language of Poynting's theorem, the total 
radiated energy equals the work done on the particles by the RR field. 
We consider a general case of arbitrary number of particles with velocity
$\boldsymbol{\beta}c$. Using Li\'{e}nard-Wiechert potential, the angular radiated
power at retarded time $t^\prime$ is \cite{maggiore2023modern}
\begin{equation}
        \dv{P(t^\prime)}{\Omega} =  \frac{\kappa}{16\pi^2 \epsilon_0 c^3}
        \left|\int\frac{\dd{\omega}}{2\pi}\omega 
        e^{-i\omega\left[t^\prime-(r/c)\right]}\tilde{\boldsymbol{J}}_\bot
        (\omega,\omega \hat{\boldsymbol{n}}/c)\right|^2.
    \label{eq14}
\end{equation}
where we defined the Doppler factor as $\kappa = \dd{t}/\dd{t^\prime} = 
1-\hat{\boldsymbol{n}}\cdot\boldsymbol{\beta}$ and $\hat{\boldsymbol{n}}
= \boldsymbol{R}/R$, $\boldsymbol{R}$ is the observation point at $t^\prime$,
$r=|\boldsymbol{r}_0(t^\prime)|$.

Assume the current density is $\boldsymbol{J} = ec\sum_i \boldsymbol{\beta}
\delta(\boldsymbol{r}-\boldsymbol{r}_0(t)-\boldsymbol{r}_i)$, then
the transformed transverse current density is $\tilde{\boldsymbol{J}}_\bot
(\omega,\omega \hat{\boldsymbol{n}}/c) = \sum_\nu \boldsymbol{e}_{\nu}
(\boldsymbol{e}_{\nu}\cdot\tilde{\boldsymbol{J}}(\omega,\omega \hat{\boldsymbol{n}}/c))$,
where $\boldsymbol{e}_{\nu}$ is the polarization vector perpendicular to
$\hat{\boldsymbol{n}}$.
For the case of steady-state CSR, the orbit of the motion is periodic
in the average frame of the beam, otherwise there will be entrance and 
exit transients. Since any finite length orbit can be periodically 
extended, we may assume the prescribed motion $\boldsymbol{r}_0(t)$ 
is periodic in the average frame. 
Thus, without any loss of generality we have the following Fourier series
$(\boldsymbol{\beta}\cdot\boldsymbol{e}_{\nu}) 
e^{ickt-i\boldsymbol{k}\cdot\boldsymbol{r}_0(t)} =
\sum_h A_{h\nu}^* e^{i\beta_h t}$ with $\beta_h=c\kappa(k-hk_0)$.
Taking this into Eq.\eqref{eq14} and averaged over a time period
of radiation $T=2\pi/ck_0$
\begin{equation}
    \left\langle \dv{P}{\Omega}\right\rangle = 
    \frac{e^2 c\kappa}{16\pi^2 \epsilon_0} \sum_{h\nu}\sum_{i,j}
    \left(\frac{h k_0}{\kappa}\right)^2 |A_{h\nu}|^2
    B_{h}^{i,j}
    \label{eq15}
\end{equation}
where $B_{h}^{i,j}=e^{-ihk_0\hat{\boldsymbol{n}}\cdot(\boldsymbol{r}_i-\boldsymbol{r}_j)}$
is the bunching factor of $h$th harmonic.
On the other side, we calculate the work done by the RR field on the test particle
by explicitly eliminating the surface terms in the transverse electric field
\begin{equation}
    \begin{aligned}
        \dv{W}{t} = &\int \dd[3]{\boldsymbol{r}} \boldsymbol{E}^\bot(\boldsymbol{r},t)\cdot
        \boldsymbol{J}(\boldsymbol{r},t) \\
        = & -\frac{e^2c^2}{2\epsilon_0}\sum_{ij}\int_{-\infty}^{+\infty}
        k^2\dd{k}\int\frac{\dd{\Omega}}{(2\pi)^3}\int_{-\infty}^{t}\dd{t^\prime}\\
        & \times \sum_{hh^\prime,\nu}A_{h\nu}^*A_{h^\prime\nu} e^{i\beta_h t^\prime - i\beta_{h^\prime} t}
        e^{-i\boldsymbol{k}\cdot (\boldsymbol{r}_i-\boldsymbol{r}_j)}.
    \end{aligned}
\end{equation}
To eliminate the surface terms at $t^\prime=t$, we do not integrate on the direction
$\hat{\boldsymbol{k}}\cdot(\boldsymbol{r}_i-\boldsymbol{r}_j)=0$. Subtracting
a subset of integration domain with zero measure will not affect calculation
of RR field.
Using contour integration technique, we shift the pole $\beta_h\rightarrow \beta_h-i0^+$,
integrate over past time and average the result over the period of motion
$T=2\pi/c\kappa k_0$

\begin{equation}
    \begin{aligned}
        \left\langle\frac{\dd{W}}{\dd{t}\dd{\Omega}}\right\rangle = & 
        -\frac{e^2c\kappa}{16\pi^2\epsilon_0} \sum_{h\nu}
        \sum_{\hat{\boldsymbol{k}}\cdot\boldsymbol{r}_{ij}\neq 0}
        \frac{(h k_u)^2}{\kappa^2}
        |A_{h\nu}|^2 B_{h}^{i,j}.
    \end{aligned}
\end{equation}

Neglecting the spontaneous radiation term ($i=j$) in Eq.\eqref{eq15}
we have the Poynting's theorem for RR field
\begin{equation}
    \left\langle \frac{\dd{W}}{\dd{t}\dd{\Omega}} \right\rangle =
    -\left\langle \dv{P}{\Omega}\right\rangle.
\end{equation}

\begin{figure}[t]
    \includegraphics[width=120pt]{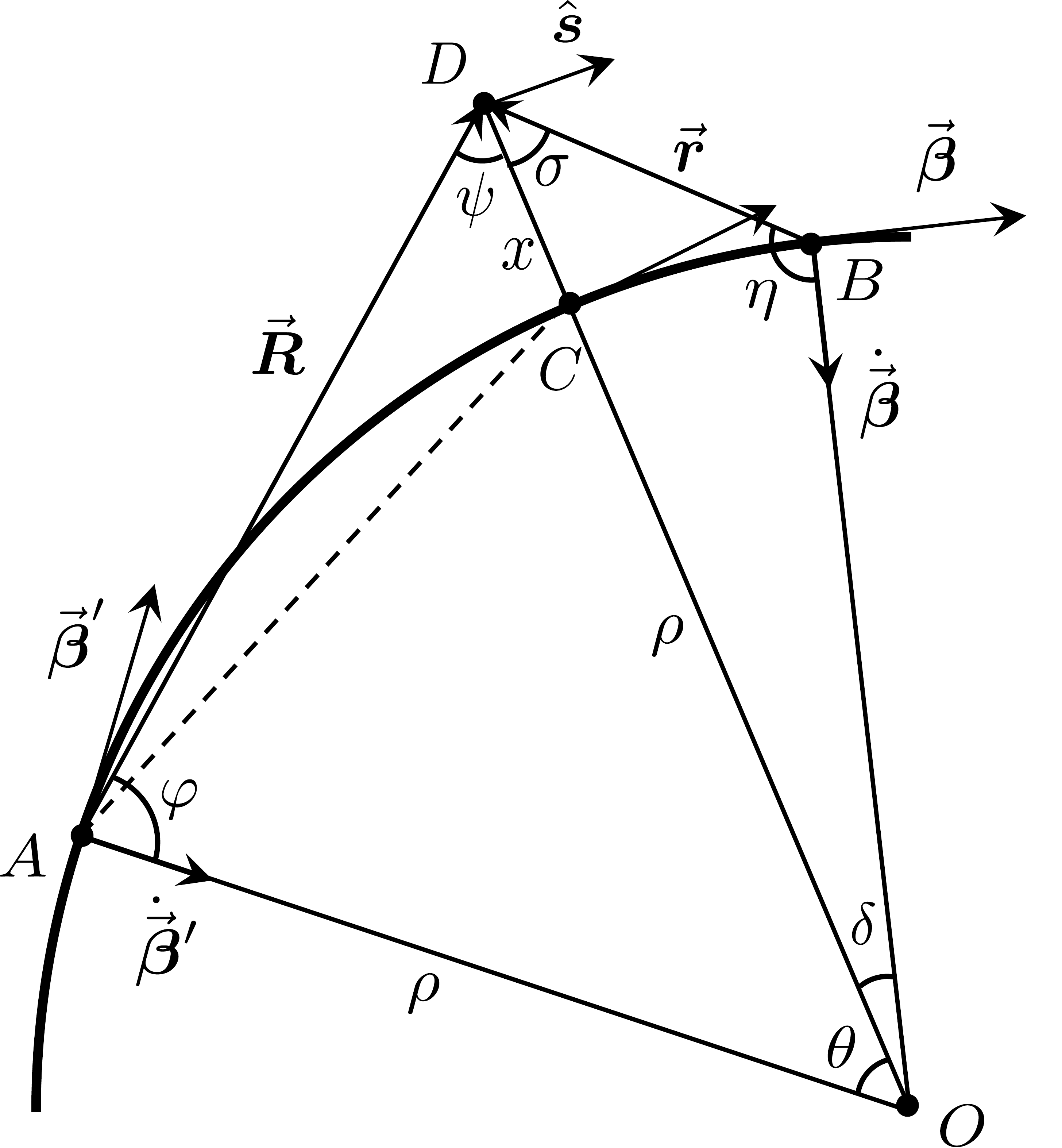}
    \caption{A source particle (A) moves along a circular orbit and
    radiates electromagnetic fields at retarded time $t^\prime$ that
    reach a test particle (D) at time $t$.}
    \label{fig2}
\end{figure}

Therefore, we verified that the defined RR field give the correct
coherent RR force.

\paragraph{Coherent synchrotron radiation.---}
Here we apply our result to analyze the CSR wake of circular orbit. In 1D
limit, our result agrees with those of Saldin and Dohlus\cite{saldin1997on,dohlus2000coherent}. 
In 2D limit, our theory gives a finite force while the others predict a diverging
force. We adopt the conventions of Cai\cite{cai2017coherent}. The definition of the
circular orbit is shown in Fig.\ref{fig2}.

Using trigonometry and define the normalized longitudinal coordinate 
$\xi = -l/2\rho$
and transverse coordinate $\chi = x/\rho$ with
$l = \beta c(t-t^\prime)-(s-s^\prime)$, where $s-s^\prime$ is the arc length between A and C.
Define the auxillary functions
 $\kappa = \sqrt{\chi^2+4(1+\chi)\sin^2\alpha}$,
$\lambda = \sqrt{1+(1+\chi)^2-2(1+\chi)\cos(\beta\kappa-2\alpha)}$, $\alpha=\theta/2$.
We find the longitudinal electric field in our theory is
\begin{widetext}
\begin{equation}
    \begin{aligned}
        E_\mathrm{rr,s} & = \frac{e}{4\pi\epsilon_0}\left\{\frac{\sin2\alpha-\beta\kappa\cos2\alpha}
        {\gamma^2\rho^2(\kappa-\beta(1+\chi)\sin2\alpha)^3} + \beta^2\frac{(\cos2\alpha-(1+\chi))((1+\chi)\sin2\alpha-\beta\kappa)}
        {\rho^2(\kappa-\beta(1+\chi)\sin2\alpha)^3}\right. \\
        & - 2\beta^2
        \frac{(3\lambda^2+1-(1+\chi)^2)\sin(\beta\kappa-2\alpha)}{\rho^2((2\lambda)^2-\beta^2(\lambda^2+1-(1+\chi)^2)^2)^{3/2}}
        + 24\beta^4\frac{(1+\chi)^2(\lambda^2+1-(1+\chi)^2)\sin^3(\beta\kappa-2\alpha)}
        {\rho^2((2\lambda)^2-\beta^2(\lambda^2+1-(1+\chi)^2)^2)^{5/2}} \\
        & \left. + \frac{8(1-\beta^2)\sin(\beta\kappa-2\alpha)}
        {\rho^2((2\lambda)^2-\beta^2(\lambda^2+1-(1+\chi)^2)^2)^{3/2}}\right\}.
    \end{aligned}
    \label{eq19}
\end{equation}
\end{widetext}

\begin{figure}[htpb]
    \includegraphics[width=240pt]{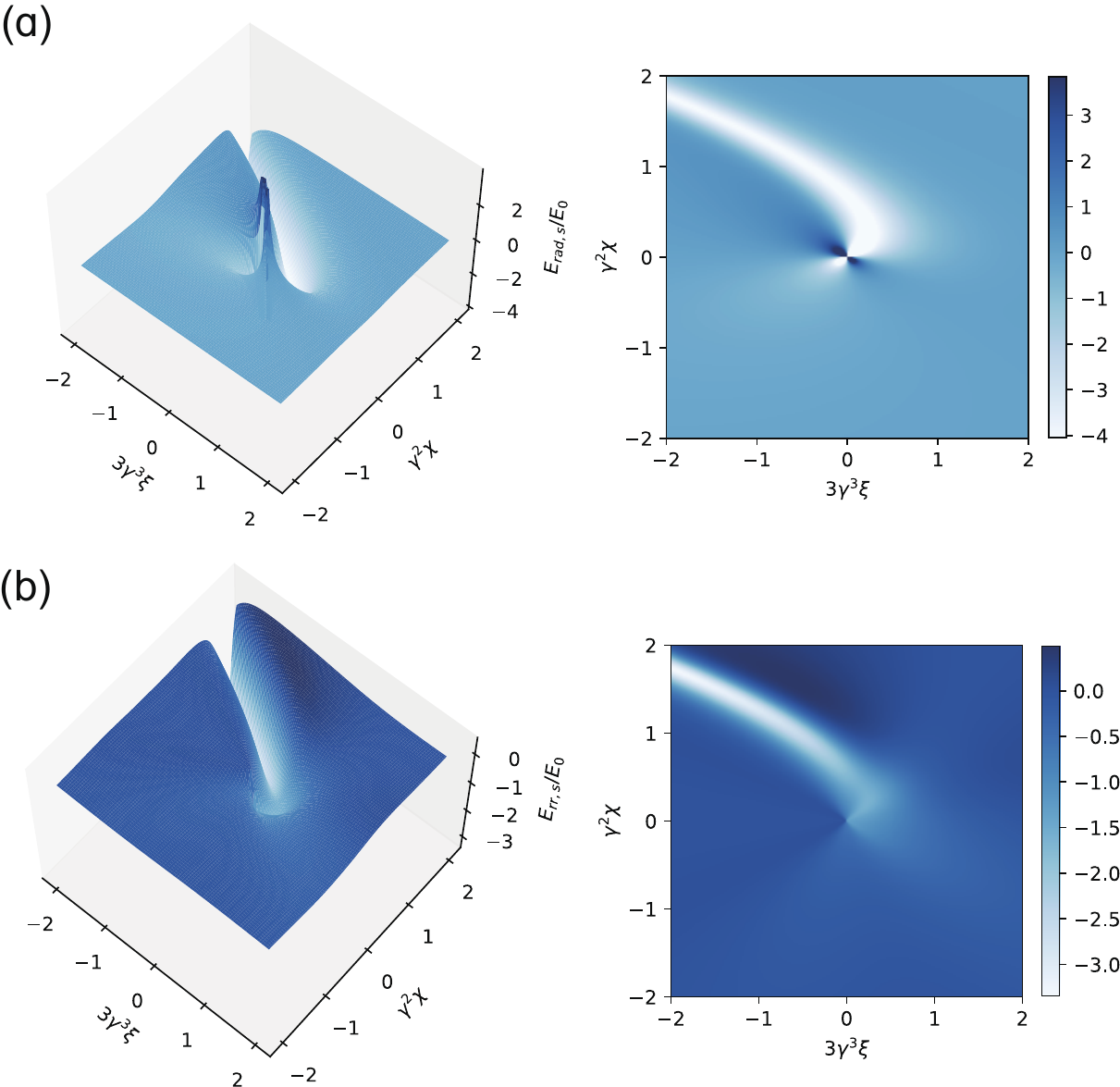}
    \caption{(a) The computed longitudinal radiation field in classical 2D
    CSR theory. (b) The computed longitudinal RR field.}
    \label{fig3}
\end{figure}

In the classical 2D CSR theory, the longitudinal field accounting for the
energy loss is the ``acceleration field'' in Li\'{e}nard-Wiechert solution.
We calculate this longitudinal field by numerically solving the retardation
condition
\begin{equation}
    \xi = \alpha-\frac{\beta}{2}\sqrt{\chi^2+4(1+\chi)\sin^2\alpha}.
\end{equation}
Assuming $\gamma=100,\rho=0.1$, the computed longitudinal field 
$E_{rad,s}$ normalized by $E_0=e\gamma^4/4\pi\epsilon_0\rho^2$ is shown
in Fig.\ref{fig3}(a), which has a quadrupole-like divergence in the center.

Again, we calculate the longitudinal RR field using Eq.\eqref{eq19},
the result normalized by $E_0$ is shown in Fig.\ref{fig3}(b). 
In contrast to the classical CSR theory, the RR field in our theory is 
everywhere finite and thus numerically verifies our claim.

\paragraph{Conclusions.---}
In this letter, we proposed a self-consistent theory describing coherent
radiation by decomposing the field into instantaneous and retarded parts.
The three terms in the decomposed fields naturally correspond to the
space-charge field, compression-induced field and radiation reaction field.
Using this theory, we described the coherent RR force and resolved
the long-standing contradictions in classical CSR theory. 

Moreover, our theory can be extended to encompass spontaneous radiation, offering
a unified description of spontaneous and coherent radiation reaction.
The derivation of this theory aligns with that of quantum electrodynamics, 
which will aid in our understanding of the classical-quantum transition. 
Therefore, our theory not only provides new insights into the physical mechanism of
coherent radiation, but will also help to investigate beam dynamics in more general
charge distributions and develop fast and accurate algorithms for simulations.

\begin{acknowledgments}
We thank Jiazhen Tang, Peng Lv of Tsinghua University for helpful discussions.
This work is supported by the National Natural Science Foundation of 
China (NSFC grant no. 11835004) and the Exploratory Program of Discipline
Development of the Department of Engineering Physics at Tsinghua 
University.
\end{acknowledgments}

\bibliography{csr}
\end{document}